\documentclass[11pt]{article}
\usepackage{amsmath,amssymb,amsfonts}
\usepackage{latexsym}   
\begin{document}
\begin{titlepage}
\begin{flushright}
CP3-07-15\\
ICMPA-MPA/2007/16\\
UTAS-PHYS-2007-07\\
June 2007\\
\end{flushright}
\begin{centering}
 
{\ }\vspace{0.5cm}
 
{\Large\bf World-line Quantisation of}

\vspace{5pt}

{\Large\bf a Reciprocally Invariant System}

\vspace{1.8cm}

Jan Govaerts\footnote{Fellow of the Stellenbosch Institute
for Advanced Study (STIAS), Stellenbosch, Republic of South Africa,
{\tt http://academic.sun.ac.za/stias/}.}$^{,}$\footnote{On sabbatical leave from the Center for Particle Physics and Phenomenology (CP3),
Institut de Physique Nucl\'eaire, Universit\'e catholique de Louvain (U.C.L.),
2, Chemin du Cyclotron, B-1348 Louvain-la-Neuve, Belgium,
E-mail: {\tt Jan.Govaerts@fynu.ucl.ac.be}.}

\vspace{0.3cm}

{\em Institute of Theoretical Physics}\\
{\em Department of Physics, University of Stellenbosch}\\
{\em Stellenbosch 7600, Republic of South Africa}\\

\vspace{0.3cm}

{\em UNESCO International Chair in Mathematical Physics and 
Applications (ICMPA)}\\
{\em University of Abomey-Calavi}\\
{\em 072 B.P. 50, Cotonou, Republic of Benin}\\

\vspace{0.7cm}

Peter D. Jarvis\footnote{Alexander von Humboldt Fellow} and
Stuart O. Morgan\footnote{Australian Postgraduate Award}

\vspace{0.3cm}

{\em School of Mathematics and Physics}\\
{\em University of Tasmania, G.P.O. Box 252C}\\
{\em 7001 Hobart, Tasmania, Australia}\\
{\em E-mail: {\tt Peter.Jarvis@utas.edu.au}, {\tt Stuart.Morgan@utas.edu.au}}

\vspace{0.7cm}

Stephen G. Low

\vspace{0.3cm}

{\em Austin, Texas, USA}\\
{\em E-mail: {\tt Stephen.Low@alumni.utexas.net}}

\clearpage

\begin{abstract}
\noindent
We present the world-line quantisation of a system invariant under the symmetries of reciprocal relativity
(pseudo-unitary transformations on ``phase space coordinates" $(x^\mu(\tau),p^\mu(\tau))$ which
preserve the Minkowski metric and the symplectic form, and global shifts in these coordinates, together
with coordinate dependent transformations of an additional compact phase coordinate, $\theta(\tau)$).
The action is that of free motion over the corresponding Weyl-Heisenberg group.
Imposition of the first class constraint, the generator of local time reparametrisations, on physical
states enforces identification of the world-line cosmological constant with a fixed value of the quadratic
Casimir of the quaplectic symmetry group $Q(D-1,1)\cong U(D-1,1)\ltimes H(D)$, the
semi-direct product of the pseudo-unitary group with the Weyl-Heisenberg group (the central extension of
the global translation group, with central extension associated to the phase variable $\theta(\tau)$).
The spacetime spectrum of physical states is identified. Even though for an appropriate range of values
the restriction enforced by the cosmological constant projects out negative norm states from
the physical spectrum, leaving over spin zero states only, the mass-squared spectrum is continuous
over the entire real line and thus includes a tachyonic branch as well.

\end{abstract}

\vspace{10pt}

\end{centering} 

\vspace{125pt}

\end{titlepage}

\setcounter{footnote}{0}

\section{Introduction}

Born reciprocity \cite{Born1949} is based on the observation of the apparent exchangeability of ``position"
and ``momentum" in much of the formalism of classical and quantum physics, and seeks to elevate this
equivalence to a fundamental principle. The idea of Born \cite{Born1949,Born1949a} and Green \cite{Green1949,Green1949a}
was to formalise this by extending the Minkowski metric of Einstein's special relativity to an invariant
metric on ``phase space coordinates"
\begin{equation}
d\ell^2=ds^2+\frac{c^4}{b^2}dm^2=
dx^\mu dx_\mu+\frac{c^2}{b^2}dp^\mu dp_\mu,
\end{equation}
where
\begin{equation}
x^\mu=(ct,\vec{x}\,),\quad p^\mu=\left(\frac{E}{c},\vec{p}\right),\quad
\eta_{\mu\nu}={\rm diag}\,(-++\cdots +),\quad \mu=0,1,2,\cdots,D-1,
\end{equation}
which can be seen as introducing a new fundamental constant, here a maximal universal unit of force $b>0$
(which can also be thought of in terms of fundamental constants of acceleration, or length, or time,
depending on the interpretation). Born and Green sought reciprocally invariant ``master" equations whose
zeroes were interpreted via multi-mass relativistic wave equations for the meson spectrum. The idea of
reciprocity has found resonance with various attempts to generalise the framework for the fundamental
interactions -- for example, in the guise of bi-crossproduct algebras and physics at the Planck scale \cite{Majid1993},
``two-time" formulations \cite{Bars}, or {\it ad hoc\/} ``noncommutative geometry" extensions of
perturbative field theory \cite{NCQFT}.

Born-Green reciprocity can be viewed \cite{Low2002,Low2005a,Low2005b,Jarvis} as an alternative paradigm for generalised wave equations, which
specify unitary irreducible representations of the full symmetry group, in the same way that relativistic wave
equations establish unitary irreducible representations of the Poincar\'e group in four dimensions. It can be
argued \cite{Low2002,Low2005a,Low2005b,Jarvis} that the appropriate invariance group is the so-called quaplectic group
$Q(3,1)\cong U(3,1)\ltimes H(4)$, or more generally in $D$ spacetime dimensions,
the group $Q(D-1,1)\cong U(D-1,1)\ltimes H(D)$ of reciprocal relativity,
the semi-direct product of the pseudo-unitary group of linear transformations
between $x^\mu$ and $p^\mu$ which preserve both the extended metric $d\ell^2$ and the symplectic form,
with the Weyl-Heisenberg group. The Wigner-Mackey method of induced representations can be applied for this
case, and $b\rightarrow\infty$ contraction limits of the appropriate generalised reciprocally invariant wave equations
should collapse to the standard relativistic wave equations of particle physics, as, for example, solutions of
the massive Klein-Gordon equation can be seen as going over to the Galilean invariant nonrelativistic wave functions
in the $c\rightarrow\infty$ limit \cite{Saletan}.

In this paper we study the alternative route to particle equations of motion via Hamiltonian quantisation of
constrained systems on the world-line \cite{Govbook}. We present the world-line quantisation of a system invariant
under the symmetries of Born-Green reciprocity, realised as transformations on ``phase space coordinates"
$(x^\mu(\tau),p^\mu(\tau))$ on the world-line, for an action, considered in Sect.~\ref{Sec2},
which is that of free motion on the associated Weyl-Heisenberg group, a guarantee from the outset for full
quaplectic invariance. These transformations are global Lorentz variations, $x^\mu\rightarrow x^\mu+{\omega^\mu}_\nu x^\nu$,
$p^\mu\rightarrow p^\mu+{\omega^\mu}_\nu p^\nu$ (in infinitesimal form, $\omega_{\mu\nu}=-\omega_{\nu\mu}$),
global quaplectic ``boosts", $x^\mu\rightarrow x^\mu+{\alpha^\mu}_\nu p^\nu/b$,
$p^\mu\rightarrow p^\mu-b{\alpha^\mu}_\nu x^\nu$ (in infinitesimal form, $\alpha_{\mu\nu}=\alpha_{\nu\mu}$),
comprising the pseudo-unitary group $U(D-1,1)$, together with global translations in both variables $x^\mu$
and $p^\mu$. The action contains terms quadratic in velocities, and also linearly coupled terms, in such a
way that both the extended metric $d\ell^2/d\tau^2$, and the symplectic form
$p_\mu(\tau)dx^\mu(\tau)/d\tau-x_\mu(\tau)dp^\mu(\tau)/d\tau$, are evident. The latter occurs in a term which
plays the role of a minimal Landau-type coupling in the kinetic energy of a further scalar variable $\theta(\tau)$,
interpreted as a phase ($S^1$) degree of freedom associated to the unit operator of the Weyl-Heisenberg algebra.
In this way global translations in $x^\mu$ and $p^\mu$, which do not leave the symplectic form invariant, are
compensated by appropriate $x$- and $p$-dependent transformations of the phase $\theta$, so that translation
invariance is restored overall. The full algebra of Noether charges thus comprises all of the conserved charges
which generate these transformations. These include the generator, ${\cal P}^\mu$, of spacetime translations
in $x^\mu$ to be interpreted as the conserved energy-momentum of the system, as well as the generator,
$L_T^{\mu\nu}$, of spacetime Lorentz transformations in $x^\mu$ and $p^\mu$, to be interpreted as the
total relativistic angular-momentum of the system. Both these quantities generate the Poincar\'e invariance
of the system, of which the representation realised by the space of quantum states determines the mass and
spin content of the quantised dynamics. Whether at the classical or quantum level the conserved charges for translations
in $x^\mu$ and $p^\mu$, ${\cal P}^\mu$ and ${\cal X}^\mu$, do not commute, but their algebra possesses a central
extension given by the conserved generator $\Pi_\theta$ of translations in the phase $\theta(\tau)$.
In any of the superselection sectors with a fixed nonvanishing discrete eigenvalue of this generator (given the
compactness of the $\theta$ part of configuration space), ${\cal P}^\mu$ and ${\cal X}^\mu$ thus fulfill a
Heisenberg algebra, and can be identified with the physical energy-momentum and possibly even the spacetime
position, respectively. The remaining Noether charges are the sum of quadratic combinations of these charges,
plus quadratic combinations of non-conserved auxiliary Heisenberg algebra generators, $\mathbb{X}^\mu$ and $\mathbb{P}^\mu$,
which commute with ${\cal P}^\mu$ and ${\cal X}^\mu$. The full symmetry algebra is thus the semi-direct product of the
conserved Heisenberg algebra, and the homogeneous charges, which indeed together generate the Lie algebra of the
quaplectic group, $Q(D-1,1)\cong U(D-1,1)\ltimes H(D)$. These considerations are detailed in Sect. \ref{Sec3}.

In Sect. \ref{Sec4} the model is extended to include local time reparametrisations. The associated Noether symmetry,
the generator of local world-line gauge transformations, is the extended Hamiltonian and becomes a first class
constraint. Following the Dirac quantisation procedure, imposition of the first class constraint on physical
states enforces the condition that the cosmological constant term, allowed by the reparametrisation invariant coupling
of the system to the world-line metric, must be identified with a discrete fixed eigenvalue of the quadratic Casimir invariant
of the quaplectic algebra, in line with well understood features of Hamiltonian quantisation \cite{Gov2}. The
decomposition of such irreducible representations of the full quaplectic group with respect to the physical
Poincar\'e group $IO(D-1,1)\cong O(D-1,1)\ltimes T(D)$ is discussed, where the generators ${\cal P}^\mu$ are
identified with the standard energy-momentum operators, the generators of spacetime translations. From this
analysis follows the spectrum of physical states classified according to the eigenspectrum of the Lorentz
covariant mass-squared and generalised Pauli-Lubanski operators. A noteworthy result is that an appropriate
choice of cosmological constant projects out any negative norm state from the physical spectrum. Nevertheless
the mass spectrum is continuous and always contains space-like, namely tachyonic states. Conclusions and possible
extensions of this work are addressed in Sect. \ref{Sec5}, while an Appendix briefly outlines the rationale
behind the choice of action used in the next Section.

\section{Dynamics and Symmetries}
\label{Sec2}

\subsection{The action}
\label{Sec2.1}

For reasons discussed in the Appendix, the degrees of freedom of the system to be considered
are the spacetime and conjugate coordinates, $x^\mu(\tau)$ and $p^\mu(\tau)$, as well as an
angular variable, $\theta(\tau)$, taking its values in the periodic range $0\le\theta\le 2\pi$.
Spacetime indices $\mu,\nu=0,1,2,\cdots,D-1$ are raised and lowered using the
Minkowski metric $\eta_{\mu\nu}={\rm diag}\,(-++\cdots+)$. The inner product defined by that
metric is denoted with a dot, {\it viz\/}, $x\cdot p=x^\mu p_\mu$.

The dynamics of the configuration space variables $(x^\mu(\tau),p^\mu(\tau),\theta(\tau))$ is
taken to follow from the action principle 
\begin{equation}
S[x^\mu,p^\mu,\theta]=\int\,d\tau\,L,\qquad
L=\frac{1}{2N_0}\left[\dot{x}^2+\kappa_0 \dot{p}^2\right]+
\frac{1}{2N_0}\frac{\alpha\kappa_0}{\lambda_0}
\left[\dot{\theta}-\lambda_0\left(\dot{x}\cdot p-x\cdot\dot{p}\right)\right]^2.
\label{eq:action1}
\end{equation}
In this expression, $N_0>0$, $\kappa_0=c^2/b^2\ge 0$, and $\alpha$ and $\lambda_0$,
with $\alpha\lambda_0>0$, are normalisation factors specifying the physical
properties of the system. Their dimension is such that the action $S$ has the
dimension of $\hbar$, namely $M\cdot L^2\cdot T^{-1}$. Since we shall not work
in ``natural" units but take the time evolution variable $\tau$ to be dimensionless,
the physical dimensions of these parameters are as follows,
\begin{equation}
\left[N_0\right]=M^{-1}\cdot T,\quad
\left[\kappa_0\right]=M^{-2}\cdot T^2,\quad
\left[\lambda_0\right]=M^{-1}\cdot L^{-2}\cdot T,\quad
\left[\alpha\right]=M\cdot T^{-1}.
\end{equation}

As explained in the Appendix, this choice of action
follows from considering free motion on the Weyl-Heisenberg group associated to the
variables $(x^\mu,p^\mu)$. The rationale behind such a choice is that, on the one hand,
it readily guarantees from the outset a dynamics which is invariant under the full
quaplectic group, and on the other hand, it generalises within this context the situation
for the ordinary relativistic particle which corresponds to motion on the group of spacetime
translations coupled to a world-line metric in a reparametrisation invariant manner.
Thus a relativistic particle dynamics invariant under the reciprocal symmetry of the
quaplectic group may likewise be constructed by coupling the above action to a world-line
metric in a reparametrisation invariant manner, as done in Sect. \ref{Sec4}. In the present
Section the symmetries of the above action are described and its dynamics understood in the next Section.
The system coupled to a world-line metric may then readily be solved based on the general considerations of \cite{Gov2}.

By construction, the action (\ref{eq:action1}) possesses a series of global symmetries generated
by the corresponding Noether charges. First one has invariance under translations in the
spacetime coordinates $x^\mu$,
\begin{equation}
x^\mu(\tau)\rightarrow x^\mu(\tau)+a^\mu,\quad
p^\mu(\tau)\rightarrow p^\mu(\tau),\quad
\theta(\tau)\rightarrow \theta(\tau)-\lambda_0 a^\mu p_\mu(\tau),
\end{equation}
of which the Noether charges are denoted ${\cal P}^\mu$, since they measure the conserved
total energy-momentum of the system and indeed possess the appropriate physical dimension.
Next one has a dual symmetry, namely invariance under
translations in the dual configuration space coordinates $p^\mu$,
\begin{equation}
x^\mu(\tau)\rightarrow x^\mu(\tau),\quad
p^\mu(\tau)\rightarrow p^\mu(\tau)+k^\mu,\quad
\theta(\tau)\rightarrow\theta(\tau)+\lambda_0 k^\mu x_\mu(\tau),
\end{equation}
of which the Noether charges are denoted ${\cal X}^\mu$, having indeed the same physical
dimensions as the spacetime coordinates $x^\mu$. The charges ${\cal X}^\mu$ and ${\cal P}^\mu$
are the generators of the Heisenberg subgroup $H(D)$ of the full quaplectic symmetry
$Q(D-1,1)\cong U(D-1,1)\ltimes H(D)$.

In infinitesimal form, the Lorentz symmetry
\begin{equation}
x^\mu(\tau)\rightarrow {\Lambda^\mu}_\nu x^\nu(\tau),\quad
p^\mu(\tau)\rightarrow{\Lambda^\mu}_\nu p^\nu(\tau),\quad
\theta(\tau)\rightarrow\theta(\tau),\quad
\eta_{\rho\sigma}\,{\Lambda^\rho}_\mu\,{\Lambda^\sigma}_\nu=\eta_{\mu\nu},
\end{equation}
reduces to the transformations
\begin{equation}
x^\mu(\tau)\rightarrow x^\mu(\tau)+{\omega^\mu}_\nu x^\nu(\tau),\quad
p^\mu(\tau)\rightarrow p^\mu(\tau)+{\omega^\mu}_\nu p^\nu(\tau),\quad
\theta(\tau)\rightarrow\theta(\tau),\qquad \omega_{\nu\mu}=-\omega_{\mu\nu},
\end{equation}
with the total relativistic angular-momentum $L_T^{\mu\nu}=-L_T^{\nu\mu}$ as the conserved Noether charge.
Likewise the dual symmetry corresponds to the symplectic transformations which in infinitesimal form read
\begin{equation}
x^\mu(\tau)\rightarrow x^\mu(\tau)+\sqrt{\kappa_0}\,{\alpha^\mu}_\nu p^\nu(\tau),\quad
p^\mu(\tau)\rightarrow p^\mu(\tau)-\frac{1}{\sqrt{\kappa_0}}{\alpha^\mu}_\nu x^\nu(\tau),\quad
\theta(\tau)\rightarrow\theta(\tau),\qquad \alpha_{\nu\mu}=\alpha_{\mu\nu},
\end{equation}
and possess conserved Noether charges denoted $M^{\mu\nu}$ with $M^{\nu\mu}=M^{\mu\nu}$.
The charges $L_T^{\mu\nu}$ and $M^{\mu\nu}$ are the generators of the pseudo-unitary subgroup
$U(D-1,1)$ of the full quaplectic symmetry $Q(D-1,1)\cong U(D-1,1)\ltimes H(D)$.

Finally, invariance under translations in the angular variable $\theta(\tau)$,
\begin{equation}
x^\mu(\tau)\rightarrow x^\mu(\tau),\quad
p^\mu(\tau)\rightarrow p^\mu(\tau),\quad
\theta(\tau)\rightarrow\theta(\tau)+\theta_0,
\end{equation}
possess a Noether charge denoted $Q_\theta$.

Rather than giving here the expressions, whether within the Lagrangian or Hamiltonian formalisms, for all
these quantities and their equations of motion, in turns out to be particularly useful to introduce a
complex parametrisation for the dynamics in which the pairs of variables $(x^\mu,p^\mu)$ for each spacetime
index $\mu=0,1,2,\cdots,D-1$ are combined into a single complex valued quantity. This choice
of representation of the dynamics is also perfectly adapted to its inherent quaplectic symmetry properties.

\subsection{The complex parametrisation}
\label{sect2.2}

By introducing the complex parametrisation of configuration space in its spacetime sector
\begin{equation}
z^\mu=\frac{1}{\sqrt{2}}\left[x^\mu+i\sqrt{\kappa_0}p^\mu\right],
\end{equation}
the action (\ref{eq:action1}) reads (a bar on top of a quantity denotes
of course its complex conjugate)
\begin{equation}
S[z^\mu,\overline{z}^\mu,\theta]=\int\,d\tau L,\qquad
L=\frac{1}{N_0}\dot{\overline{z}}\cdot\dot{z}+\frac{1}{2N_0}\frac{\alpha\kappa_0}{\lambda_0}
\left[\dot{\theta}+i\frac{\lambda_0}{\sqrt{\kappa_0}}\left(\dot{\overline{z}}\cdot z
-\overline{z}\cdot\dot{z}\right)\right]^2.
\label{eq:action2}
\end{equation}

The previously discussed global symmetries are then expressed as,
\begin{equation}
z^\mu(\tau)\rightarrow z^\mu(\tau)+{\Omega^\mu}_\nu z^\nu(\tau)+ A^\mu,\quad
\theta(\tau)\rightarrow\theta(\tau)+\theta_0
-i\frac{\lambda_0}{\sqrt{\kappa_0}}
\left[\overline{z}(\tau)\cdot A-z(\tau)\cdot\overline{A}\right],
\end{equation}
with the correspondences
\begin{equation}
\Omega_{\mu\nu}=\omega_{\mu\nu}-i\alpha_{\mu\nu},\quad
\overline{\Omega}_{\mu\nu}=-\Omega_{\nu\mu},\quad
A^\mu=\frac{1}{\sqrt{2}}\left[a^\mu+i\sqrt{\kappa_0}\,k^\mu\right].
\end{equation}
In this form it is clear that the global symmetry group of the system,
namely the so-called quaplectic group \cite{Low2002,Low2005a,Low2005b}, is indeed isomorphic to
$Q(D-1,1)\cong U(D-1,1)\ltimes H(D)$. The associated Noether charges are to be
denoted by, respectively for the symmetry parameters $A_\mu$, $\Omega_{\mu\nu}$ and $\theta_0$,
\begin{equation}
Q^\mu,\qquad Q^{\mu\nu}=-\overline{Q}^{\nu\mu},\qquad Q_\theta,
\end{equation}
with the following correspondences with the previous notations,
\begin{equation}
Q^\mu=\frac{1}{\sqrt{2}}\left[{\cal P}^\mu-\frac{i}{\sqrt{\kappa_0}}{\cal X}^\mu\right],\quad
Q^{\mu\nu}=\frac{1}{2}L_T^{\mu\nu}\,+\,\frac{1}{2}i\,M^{\mu\nu}.
\end{equation}

\section{Hamiltonian Formulation and Canonical Quantisation}
\label{Sec3}

\subsection{The Noether algebra}
\label{Sec3.1}

Within the Hamiltonian formulation of the dynamics, momenta canonically conjugate to the configuration variables
$x^\mu$, $p^\mu$, $z^\mu$ and $\theta$ are denoted, respectively, by $\Pi^\mu_x$, $\Pi^\mu_p$,
$\Pi^\mu$ and $\Pi_\theta$, with their canonical equal time Poisson brackets. Henceforth we shall already
consider the canonically quantised system, in which these variables are operators of which the commutation
relations are given by the result of the corresponding Poisson brackets multiplied by $i\hbar$ (we shall
refrain from introducing a notation distinguishing between operators and their classical counterparts, but
the difference should be clear from the context and be kept in mind). Hence
the quantised dynamics is realised as a representation of the following general tensor product of
Heisenberg algebras,
\begin{equation}
\left[x^\mu,\Pi^\nu_x\right]=i\hbar\,\eta^{\mu\nu},\quad
\left[p^\mu,\Pi^\nu_p\right]=i\hbar\,\eta^{\mu\nu},\quad
\left[z^\mu,\Pi^\nu\right]=i\hbar\eta^{\mu\nu},\quad
\left[\theta,\Pi_\theta\right]=i\hbar.
\end{equation}
The correspondence between the quantities $\Pi^\mu_x$, $\Pi^\mu_p$ and $\Pi^\mu$ is such that
\begin{equation}
\Pi^\mu=\frac{1}{\sqrt{2}}\left[\Pi^\mu_x-\frac{i}{\sqrt{\kappa_0}}\Pi^\mu_p\right].
\end{equation}
Note also that as operators, a quantity with a bar on top stands for the adjoint of that operator,
{\it viz\/}, $\overline{z}^\mu={z^\mu}^\dagger$, with respect to the implicit inner product on the
space of quantum states for which the basic operators $x^\mu$, $p^\mu$, $\theta$ and their
conjugate momenta $\Pi^\mu_x$, $\Pi^\mu_p$ and $\Pi_\theta$ are hermitian and self-adjoint.

The identification of the previously discussed Noether charges in terms of these quantities
is readily achieved, leading to the expressions,
\begin{equation}
Q^\mu=\Pi^\mu-i\frac{\lambda_0\Pi_\theta}{\sqrt{\kappa_0}}\,\overline{z}^\mu,\quad
Q^{\mu\nu}=z^\mu\,\Pi^\nu\,-\,\overline{z}^\nu\,\overline{\Pi}^\mu,\quad
Q_\theta=\Pi_\theta.
\end{equation}
Separating the real and imaginary parts of $Q^\mu$ and $Q^{\mu\nu}$, one also finds
\begin{equation}
{\cal P}^\mu=\Pi^\mu_x-\lambda_0\Pi_\theta p^\mu,\quad
{\cal X}^\mu=\Pi^\mu_p+\lambda_0\Pi_\theta x^\mu,
\end{equation}
\begin{equation}
L_T^{\mu\nu}=x^\mu\,\Pi^\nu_x-x^\nu\,\Pi^\mu_x + p^\mu\,\Pi^\nu_p - p^\nu\,\Pi^\mu_p,
\end{equation}
\begin{equation}
M^{\mu\nu}=\sqrt{\kappa_0}\left[ p^\mu\,\Pi^\nu_x+p^\nu\,\Pi^\mu_x\right]
-\frac{1}{\sqrt{\kappa_0}}\left[x^\mu\,\Pi^\nu_p + x^\nu\,\Pi^\mu_p\right].
\end{equation}

In view of these expressions, let us also introduce the dual combinations
\begin{equation}
\mathbb{Q}^\mu=\Pi^\mu+i\frac{\lambda_0\Pi_\theta}{\sqrt{\kappa_0}}\overline{z}^\mu=
\frac{1}{\sqrt{2}}\left[\mathbb{P}^\mu-\frac{i}{\sqrt{\kappa_0}}\mathbb{X}^\mu\right],
\end{equation}
with thus
\begin{equation}
\mathbb{P}^\mu=\Pi^\mu_x+\lambda_0\Pi_\theta p^\mu,\quad
\mathbb{X}^\mu=\Pi^\mu_p-\lambda_0\Pi_\theta x^\mu.
\end{equation}
Note that in terms of these variables one may write
\begin{equation}
Q^{\mu\nu}=\frac{1}{2}\left[z^\mu Q^\nu-\overline{z}^\nu\overline{Q}^\mu\right]\,+\,
\frac{1}{2}\left[ z^\mu\mathbb{Q}^\nu - \overline{z}^\nu\overline{\mathbb{Q}}^\mu\right],
\end{equation}
as well as
\begin{equation}
L_T^{\mu\nu}=L_{\rm orbital}^{\mu\nu}+\left[p^\mu\mathbb{X}^\nu-p^\nu\mathbb{X}^\mu\right],\qquad
L_{\rm orbital}^{\mu\nu}=x^\mu{\cal P}^\nu-x^\nu{\cal P}^\mu,
\end{equation}
\begin{equation}
M^{\mu\nu}=\frac{1}{2}\sqrt{\kappa_0}\left[p^\mu{\cal P}^\nu+p^\mu\mathbb{P}^\nu\right]
-\frac{1}{2\sqrt{\kappa_0}}\left[x^\mu{\cal X}^\nu+x^\mu\mathbb{X}^\nu\right]\,+\,
\left(\mu\leftrightarrow\nu\right).
\end{equation}
The above expression for the total angular-momentum in which the orbital angular-momentum
contribution $L_{\rm orbital}^{\mu\nu}$ is isolated, clearly shows that whereas
the degrees of freedom $x^\mu=(ct,\vec{x}\,)$
may be interpreted as describing the position of the reciprocal particle in Minkowski spacetime,
the dual variables $p^\mu=(E/c,\vec{p}\,)$ play in fact the r\^ole of internal degrees of freedom
which may carry some internal spin structure when properly excited. This separation in the r\^oles
played by the two types of variables $({\cal X}^\mu,{\cal P}^\mu)$, namely $Q^\mu$ on the one hand, and
$(\mathbb{X}^\mu,\mathbb{P}^\mu)$, namely $\mathbb{Q}^\mu$ on the other hand, is to be exploited further later on.
Incidentally, one may also write, provided however $\lambda_0\Pi_\theta\ne 0$,
\begin{equation}
Q^{\mu\nu}=-i\frac{\sqrt{\kappa_0}}{2\lambda_0\Pi_\theta}
\left[\overline{Q}^\mu Q^\nu\,-\,\overline{\mathbb{Q}}^\mu\mathbb{Q}^\nu\right],
\end{equation}
or in real form,
\begin{equation}
L_T^{\mu\nu}=\frac{1}{2\lambda_0\Pi_\theta}\left[\left({\cal X}^\mu{\cal P}^\nu-{\cal X}^\nu{\cal P}^\mu\right)\,-\,
\left(\mathbb {X}^\mu\mathbb{P}^\nu-\mathbb{X}^\nu\mathbb{P}^\mu\right)\right],
\end{equation}
\begin{equation}
M^{\mu\nu}=-\frac{1}{2\lambda_0\Pi_\theta}\frac{1}{\sqrt{\kappa_0}}
\left[\left({\cal X}^\mu{\cal X}^\nu+\kappa_0{\cal P}^\mu{\cal P}^\nu\right)\,-\,
\left(\mathbb{X}^\mu\mathbb{X}^\nu+\kappa_0\mathbb{P}^\mu\mathbb{P}^\nu\right)\right],
\end{equation}
results which once again display the dual r\^oles played by the two sectors $Q^\mu$ and $\mathbb{Q}^\mu$
which, together with $(\theta,\Pi_\theta)$, provide an alternative parametrisation of the phase space of the
system, namely through the change of variables
$(x^\mu,\Pi^\mu_x;p^\mu,\Pi^\mu_p;\theta,\Pi_\theta)\leftrightarrow(Q^\mu;\mathbb{Q}^\mu;\theta,\Pi_\theta)$.

All these considerations having been made explicit, the evaluation of the algebra of commutation relations
for all Noether charges is straightforward. The nonvanishing commutators are found to be,
\begin{equation}
\left[Q^\mu,\overline{Q}^\nu\right]=2\hbar\,\frac{\lambda_0\Pi_\theta}{\sqrt{\kappa_0}}\,\eta^{\mu\nu},
\end{equation}
\begin{equation}
\left[Q^{\mu\nu},Q^\rho\right]=i\hbar\,\eta^{\mu\rho}\,Q^\nu,\qquad
\left[Q^{\mu\nu},\overline{Q}^\rho\right]=-i\hbar\,\eta^{\nu\rho}\,\overline{Q}^\mu,
\end{equation}
\begin{equation}
\left[Q^{\mu\nu},Q^{\rho\sigma}\right]=i\hbar\,\left[
\eta^{\mu\sigma}\,Q^{\rho\nu}\,-\,\eta^{\nu\rho}\,Q^{\mu\sigma}\right].
\end{equation}
Furthermore the operators $Q^\mu$ and $\overline{Q}^\mu$ commute with both
$\mathbb{Q}^\mu$ and $\overline{\mathbb{Q}}^\mu$, while one also has
\begin{equation}
\left[\mathbb{Q}^\mu,\overline{\mathbb{Q}}^\nu\right]=
-2\hbar\,\frac{\lambda_0\Pi_\theta}{\sqrt{\kappa_0}}\,\eta^{\mu\nu}.
\end{equation}
Written in their real form, these commutation relations also correspond to the algebra
of Noether charges,
\begin{equation}
\left[{\cal X}^\mu,{\cal P}^\nu\right]=2i\hbar\,\lambda_0\Pi_\theta\,\eta^{\mu\nu},\qquad
\left[\mathbb{X}^\mu,\mathbb{P}^\nu\right]=-2i\hbar\,\lambda_0\Pi_\theta\,\eta^{\mu\nu},
\end{equation}
\begin{equation}
\left[L_T^{\mu\nu},{\cal X}^\rho\right]=i\hbar\,\left[\eta^{\mu\rho}\,{\cal X}^\nu\,-\,\eta^{\nu\rho}\,{\cal X}^\mu\right],\qquad
\left[L_T^{\mu\nu},{\cal P}^\rho\right]=i\hbar\,\left[\eta^{\mu\rho}\,{\cal P}^\nu\,-\,\eta^{\nu\rho}\,{\cal P}^\mu\right],
\end{equation}
\begin{equation}
\left[M^{\mu\nu},{\cal X}^\rho\right]=i\hbar\,\sqrt{\kappa_0}
\left[\eta^{\mu\rho}\,{\cal P}^\nu\,+\,\eta^{\nu\rho}\,{\cal P}^\mu\right],\qquad
\left[M^{\mu\nu},{\cal P}^\rho\right]=-i\hbar\,\frac{1}{\sqrt{\kappa_0}}
\left[\eta^{\mu\rho}\,{\cal X}^\nu\,+\,\eta^{\nu\rho}\,{\cal X}^\mu\right],
\end{equation}
\begin{equation}
\left[L_T^{\mu\nu},L_T^{\rho\sigma}\right]=i\hbar\,\left[
\eta^{\mu\rho}\,L_T^{\nu\sigma}-\eta^{\mu\sigma}\,L_T^{\nu\rho}
-\eta^{\nu\rho}\,L_T^{\mu\sigma}+\eta^{\nu\sigma}\,L_T^{\mu\rho}\right],
\end{equation}
\begin{equation}
\left[L_T^{\mu\nu},M^{\rho\sigma}\right]=i\hbar\,\left[
\eta^{\mu\rho}\,M^{\nu\sigma}+\eta^{\mu\sigma}\,M^{\nu\rho}
-\eta^{\nu\rho}\,M^{\mu\sigma}-\eta^{\nu\sigma}\,M^{\mu\rho}\right],
\end{equation}
\begin{equation}
\left[M^{\mu\nu},M^{\rho\sigma}\right]=i\hbar\,\left[
\eta^{\mu\rho}\,L_T^{\nu\sigma}+\eta^{\mu\sigma}\,L_T^{\nu\rho}
+\eta^{\nu\rho}\,L_T^{\mu\sigma}+\eta^{\nu\sigma}\,L_T^{\mu\rho}\right].
\end{equation}

Note that the $U(D-1,1)$ algebra generated by $Q^{\mu\nu}$ possesses the linear Casimir 
\begin{equation}
C_1=-i{Q^\mu}_\mu=\frac{1}{2}{M^\mu}_\mu=
\sqrt{\kappa_0}\,p\cdot\Pi_x\,-\,\frac{1}{\sqrt{\kappa_0}}\,x\cdot\Pi_p,\qquad
\overline{C}_1=C_1.
\end{equation}
One has
\begin{equation}
\left[C_1,Q^\mu\right]=\hbar\,Q^\mu,\qquad
\left[C_1,\overline{Q}^\mu\right]=-\hbar\,\overline{Q}^\mu,\qquad
\left[C_1,Q^{\mu\nu}\right]=0,
\end{equation}
or in real form,
\begin{equation}
\left[C_1,{\cal X}^\mu\right]=i\hbar\sqrt{\kappa_0}\,{\cal P}^\mu,\qquad
\left[C_1,{\cal P}^\mu\right]=-i\frac{\hbar}{\sqrt{\kappa_0}}\,{\cal X}^\mu,\qquad
\left[C_1,L_T^{\mu\nu}\right]=0,\qquad
\left[C_1,M^{\mu\nu}\right]=0.
\end{equation}

Another global symmetry of the dynamics has yet to be addressed, namely its invariance
under constant translations in the time evolution parameter $\tau$, $\tau\rightarrow\tau+\tau_0$,
of which the conserved Noether charge is the canonical Hamiltonian $H$. A straightforward
evaluation of this quantity finds the following identification,
\begin{equation}
H=\frac{1}{2}N_0\left[\overline{\mathbb{Q}}\cdot\mathbb{Q}+
\mathbb{Q}\cdot\overline{\mathbb{Q}}\,+\,\frac{\lambda_0}{\alpha\kappa_0}\Pi^2_\theta\right].
\end{equation}
Since the sectors $Q^\mu$ and $\mathbb{Q}^\mu$ commute with one another, this form of the
Hamiltonian makes it explicit that indeed the Noether charges $Q^\mu$ are conserved. That
the Noether charges $Q^{\mu\nu}=z^\mu\Pi^\nu-\overline{z}^\nu\overline{\Pi}^\mu$ are conserved
readily follows from a simple direct calculation. In other words, the commutators of all Noether
charges with the Hamiltonian do indeed vanish. 

As a matter of fact, there exists an alternative representation of the Hamiltonian operator
involving the $U(D-1,1)$ linear Casimir $C_1$. Expressing $\mathbb{Q}^\mu$ in terms of $Q^\mu$, one finds
\begin{equation}
H=\frac{1}{2}N_0\left[\overline{Q}\cdot Q+ Q\cdot\overline{Q}
+4\frac{\lambda_0\Pi_\theta}{\sqrt{\kappa_0}}\,C_1\,
+\frac{\lambda_0}{\alpha\kappa_0}\Pi_\theta^2\right].
\end{equation}
One also has
\begin{equation}
\overline{Q}\cdot Q+Q\cdot\overline{Q}={\cal P}^2+\frac{1}{\kappa_0}{\cal X}^2,\qquad
\overline{\mathbb{Q}}\cdot\mathbb{Q}+\mathbb{Q}\cdot\overline{\mathbb{Q}}=
\mathbb{P}^2+\frac{1}{\kappa_0}\mathbb{X}^2.
\end{equation}
Incidentally, these results imply that the quadratic Casimir operator of the quaplectic
algebra $Q(D-1,1)\cong U(D-1,1)\ltimes H(D)$ is given by
\begin{equation}
C_2=\overline{\mathbb{Q}}\cdot\mathbb{Q}+\mathbb{Q}\cdot\overline{\mathbb{Q}},
\end{equation}
with the following relation to the linear $U(D-1,1)$ Casimir $C_1$,
\begin{equation}
C_2=\overline{Q}\cdot Q+Q\cdot\overline{Q}+4\frac{\lambda_0\Pi_\theta}{\sqrt{\kappa_0}}\,C_1=
{\cal P}^2+\frac{1}{\kappa_0}{\cal X}^2+4\frac{\lambda_0\Pi_\theta}{\sqrt{\kappa_0}}\,C_1.
\end{equation}

It is worth noting that the combinations of variables $(x^\mu,\Pi^\mu_x,p^\mu,\Pi^\mu_p)$
to form the quantities $({\cal X}^\mu,{\cal P}^\mu,\mathbb{X}^\mu,\mathbb{P}^\mu)$, and
finally $(Q^\mu,\mathbb{Q}^\mu)$, with in particular the Hamiltonian solely expressed
in terms of $\mathbb{Q}^\mu$ while the sectors $Q^\mu$ and $\mathbb{Q}^\mu$ are commuting
with each defining a Heisenberg algebra, is very much reminiscent of
the ordinary Landau problem. What corresponds to the Euclidean plane coordinates $(x,y)$ and the
magnetic field in the latter case are now, respectively, the pairs of variables $(x^\mu,p^\mu)$ for
each of the spacetime components $\mu=0,1,2,\cdots,D-1$, and the conserved Noether charge $Q_\theta=\Pi_\theta$, except
for an overall minus multiplying the contribution of the time component sector $\mu=0$
to the total Hamiltonian. Indeed, $\Pi_\theta$ is conserved and commutes with all operators contributing to
the quantum dynamics. This remark allows one to consider now the
diagonalisation problem of both the Hamiltonian $H$ and the total energy-momentum
${\cal P}^\mu$ of the system.

\subsection{Quantum spectrum: the generic situation $\Pi_\theta\ne 0$}
\label{Sec3.2}

Since $\Pi_\theta$ is conserved and commutes with all other operators (except $\theta$ of course), it is
most convenient to consider the diagonalisation problem in each of the superselection sectors defined by
each of the discrete $\Pi_\theta$ eigenstates $|n\rangle$,
\begin{equation}
\Pi_\theta|n\rangle=\hbar(n+\lambda)|n\rangle,\qquad n\in\mathbb{Z},\quad \lambda\in[0,1[,
\end{equation}
where $\lambda$ (defined modulo 1) parametrises a U(1) holonomy and the freedom in the choice
of representation for the Heisenberg algebra $[\theta,\Pi_\theta]=i\hbar$ associated to the compact degree
of freedom $0\le\theta\le 2\pi$ \cite{Gov3}. In order to exploit the noted analogy with the ordinary
Landau problem, let us consider any given superselection sector associated to such a nonvanishing
eigenvalue, $\Pi_\theta=\hbar(n+\lambda)\ne 0$. This is guaranteed for all $n\in\mathbb{Z}$ provided $\lambda\ne 0$,
or else if $\lambda=0$ when $n\ne 0$. The particular situation when $\Pi_\theta=0$ is to be
considered separately in Sect. \ref{Sec3.3}.

The $\Pi_\theta$ superselection sector having been specified in this manner, for what
concerns the remaining variables $(x^\mu,\Pi^\mu_x;p^\mu,\Pi^\mu_p)$ let us introduce
the Fock algebra generators
\begin{equation}
a^\mu_+=\sqrt{\frac{|\lambda_0\Pi_\theta|}{2\hbar\sqrt{\kappa_0}}}
\left[\overline{z}^\mu+i\frac{\sqrt{\kappa_0}}{|\lambda_0\Pi_\theta|}\Pi^\mu\right],\qquad
{a^\mu_+}^\dagger=\sqrt{\frac{|\lambda_0\Pi_\theta|}{2\hbar\sqrt{\kappa_0}}}
\left[z^\mu-i\frac{\sqrt{\kappa_0}}{|\lambda_0\Pi_\theta|}\overline{\Pi}^\mu\right],
\end{equation}
\begin{equation}
a^\mu_-=\sqrt{\frac{|\lambda_0\Pi_\theta|}{2\hbar\sqrt{\kappa_0}}}
\left[z^\mu+i\frac{\sqrt{\kappa_0}}{|\lambda_0\Pi_\theta|}\overline{\Pi}^\mu\right],\qquad
{a^\mu_-}^\dagger=\sqrt{\frac{|\lambda_0\Pi_\theta|}{2\hbar\sqrt{\kappa_0}}}
\left[\overline{z}^\mu
-i\frac{\sqrt{\kappa_0}}{|\lambda_0\Pi_\theta|}\Pi^\mu\right].
\end{equation}
These operators define two commuting sets of Fock algebras, for each of the
spacetime components $\mu,\nu=0,1,2,\cdots,D-1$,
\begin{equation}
\left[a^\mu_s,{a^\nu_{s'}}^\dagger\right]=\delta_{s,s'}\,\eta^{\mu\nu},\qquad s,s'=+,-.
\end{equation}

As a function of the sign of the product $\lambda_0\Pi_\theta$,
$\eta={\rm sign}\,\left(\lambda_0\Pi_\theta\right)$, we then have
\begin{equation}
{\rm If}\ \eta=+1:\qquad
Q^\mu=\sqrt{\frac{2\hbar|\lambda_0\Pi_\theta|}{\sqrt{\kappa_0}}}\,\left(-i\,a^\mu_+\right),\qquad
\mathbb{Q}^\mu=\sqrt{\frac{2\hbar|\lambda_0\Pi_\theta|}{\sqrt{\kappa_0}}}\,\left(i\,{a^\mu_-}^\dagger\right);
\end{equation}
\begin{equation}
{\rm If}\ \eta=-1:\qquad
Q^\mu=\sqrt{\frac{2\hbar|\lambda_0\Pi_\theta|}{\sqrt{\kappa_0}}}\,\left(i\,{a^\mu_-}^\dagger\right),\qquad
\mathbb{Q}^\mu=\sqrt{\frac{2\hbar|\lambda_0\Pi_\theta|}{\sqrt{\kappa_0}}}\,\left(-i\,a^\mu_+\right),
\end{equation}
so that,
\begin{equation}
\Pi^\mu=\frac{1}{2}\left(Q^\mu+\mathbb{Q}^\mu\right)=
i\sqrt{\frac{\hbar|\lambda_0\Pi_\theta|}{2\sqrt{\kappa_0}}}
\left[{a^\mu_-}^\dagger\,-\,a^\mu_+\right],
\end{equation}
and
\begin{equation}
{\cal X}^\mu=\eta\sqrt{\hbar|\lambda_0\Pi_\theta|\sqrt{\kappa_0}}\left[{a^\mu_\eta}^\dagger+a^\mu_\eta\right],\qquad
{\cal P}^\mu=i\sqrt{\frac{\hbar|\lambda_0\Pi_\theta|}{\sqrt{\kappa_0}}}
\left[{a^\mu_\eta}^\dagger-a^\mu_\eta\right],
\end{equation}
while the total angular-momentum reduces to
\begin{equation}
L_T^{\mu\nu}=\frac{1}{2\lambda_0\Pi_\theta}\left[{\cal X}^\mu{\cal P}^\nu\,-\,
{\cal X}^\nu{\cal P}^\mu\right]\,-\,i\hbar\left[{a^\mu_{-\eta}}^\dagger a^\nu_{-\eta}\,-\,
{a^\nu_{-\eta}}^\dagger a^\mu_{-\eta}\right].
\label{eq:Ltot}
\end{equation}
Finally the Hamiltonian is diagonalised in the form
\begin{equation}
H=2\hbar N_0\frac{|\lambda_0\Pi_\theta|}{\sqrt{\kappa_0}}\,a^\dagger_{-\eta}\cdot a_{-\eta}\ +\
\frac{1}{2}N_0\frac{\lambda_0}{\alpha\kappa_0}\,\Pi^2_\theta\ +\
\hbar N_0\frac{|\lambda_0\Pi_\theta|}{\sqrt{\kappa_0}}\,D.
\end{equation}

Note that the expression (\ref{eq:Ltot}) shows that the generalised Pauli-Lubanski operator
\begin{equation}
W_{\mu_1\mu_2\cdots\mu_{D-3}}=\frac{1}{2}\epsilon_{\mu_1\mu_2\cdots\mu_{D-3}\mu\nu\rho}\,
{\cal P}^{\mu}\,L_T^{\nu\rho},\qquad
S^{\mu\nu\rho}=\frac{(-1)^{D-3}}{(D-3)!}\,\epsilon^{\mu\nu\rho\mu_1\mu_2\cdots\mu_{D-3}}\,
W_{\mu_1\mu_2\cdots\mu_{D-3}},
\label{eq:PauliLub}
\end{equation}
which characterises the internal spin representation of quantum states, receives contributions
only from the degrees of freedom $(a^\mu_{-\eta},a^\mu_{-\eta}{}^\dagger)$. The latter are thus to be interpreted
as the internal degrees of freedom of the system, while $({\cal X}^\mu,{\cal P}^\mu)$ are the
spacetime ones commuting with the internal ones, and defining their own Heisenberg algebra,
$\left[{\cal X}^\mu,{\cal P}^\nu\right]=2i\hbar\lambda_0\Pi_\theta\eta^{\mu\nu}\ne 0$. This
interpretation is also consistent with the fact that only the internal degrees of
freedom contribute to the spectrum of the Hamiltonian generator of time evolution in $\tau$.

A complete diagonalisation, namely the identification of a complete basis of
quantum states is thus achieved. Given any of the $\Pi_\theta$
eigenstates with $\Pi_\theta\ne 0$, one takes its tensor product
with any of the ${\cal P}^\mu$ eigenstates, $|{\cal P}^\mu\rangle$, as well as
with any of the Fock states associated to the Fock algebra generated by $a^\mu_{-\eta}$ and
$a^\mu_{-\eta}{}^\dagger$. In the latter sector {\it a priori} one may have two choices
to be contemplated, each with its drawbacks. For the first choice all operators $a^\mu_{-\eta}$
are considered as annihilation operators of the normalised Fock vacuum, with all operators
$a^\mu_{-\eta}{}^\dagger$ thus being creation operators. In such a situation, the Fock vacuum is
Lorentz invariant under the action of $L_T^{\mu\nu}$, and manifest Lorentz covariance of
all quantum states is ensured throughout. However, because of the negative definite signature
of the time component of the Minkowski metric, $\eta_{00}=-1$, any state involving
an odd power of the creation operator $a^{\mu=0}_{-\eta}{}^\dagger$ is of negative norm.
The appearance of such negative norm states is generic in any Lorentz covariant quantum system
with internal degrees of freedom. Nevertheless, the total number operator contribution to the Hamiltonian,
$N=a^\dagger_{-\eta}\cdot a_{-\eta}$, remains then positive definite, with a degeneracy at
each level equal to the number of $D$-partitions of $N$ over the natural numbers, which
corresponds to the dimension of the totally $N$-symmetric representation of $SU(D)$. As an alternative
choice avoiding the appearance of these negative norm states, one may wish to use rather
$a^{\mu=0}_{-\eta}{}^\dagger$ and $a^{\mu=i}_{-\eta}$ with $i=1,2,\cdots,D-1$ as annihilation
operators of the Fock vacuum. However such a choice would break Lorentz invariance of the Fock
vacuum of the quantised system, a most unwelcome feature. It would also imply an unbounded below spectrum for
the quantum Hamiltonian $H$.

Consequently the choice to be made is the first one which is manifestly Lorentz covariant.
It is to be hoped that in a manner similar to what happens in string theory for instance, once the symmetry
under translations in the time evolution parameter $\tau$ is gauged by coupling the dynamics to
a world-line metric in a diffeomorphic invariant manner, the resulting first class constraint is
such that these negative norm states are projected out from the physical spectrum, namely that the
physical spectrum is restricted to lie within the lowest level $N=0$ of the Hamiltonian spectrum. Nonetheless,
the fact that the generators of all Poincar\'e symmetries, namely ${\cal P}^\mu$ and $L_T^{\mu\nu}$,
commute with the Hamiltonian implies that such a physical spectrum still transforms covariantly
under the full Lorentz group, resulting in a consistent physical interpretation of the system
independent of the choice of world-line parametrisation.

Given that Lorentz covariant choice of Fock states, it follows that the spectrum of the Hamiltonian is organised
into discrete levels quite analogous to the Landau levels of the ordinary Landau problem.
Each of these excitation levels distinguished by the eigenvalue of the number operator $N$
carries a two-fold degeneracy, one which is finite and corresponds to the $D$-partitions of $N$
into the natural numbers, and the other which is infinite non-countable and parametrised by the
real eigenvalue spectrum of energy-momentum values ${\cal P}^\mu$. At any such level $N=0,1,2,\cdots$,
the corresponding states are organised into irreducible representations of the full $D$-dimensional
Poincar\'e group, labelled by the representation of the generalised Pauli-Lubanski tensor
(\ref{eq:PauliLub}) generated by the operators $(a^\mu_{-\eta},a^\mu_{-\eta}{}^\dagger)$, as well as
the Lorentz invariant measuring the mass-squared value of the states, $(Mc)^2=-{\cal P}^2$.
Note that the latter may take its value anywhere in the continuous
spectrum defined by the entire real line. Thus for instance even for the lowest states with $N=0$
which are all of vanishing internal spin and of strictly positive norm, one obtains a continuous mass
spectrum with a tachyonic branch for space-like energy-momenta eigenvalues ${\cal P}^\mu$.

\subsection{Quantum spectrum: the particular case $\Pi_\theta=0$}
\label{Sec3.3}

In the $\Pi_\theta$ superselection sector with a vanishing eigenvalue $\Pi_\theta=0$, most quantities
simplify drastically,
\begin{equation}
Q^\mu=\mathbb{Q}^\mu=\Pi^\mu,\qquad
{\cal P}^\mu=\mathbb{P}^\mu=\Pi^\mu_x,\qquad
{\cal X}^\mu=\mathbb{X}^\mu=\Pi^\mu_p,
\end{equation}
so that
\begin{equation}
\left[{\cal X}^\mu,{\cal P}^\nu\right]=0,
\end{equation}
as well as
\begin{equation}
H=\frac{1}{2}N_0\left[{\cal P}^2+\frac{1}{\kappa_0}{\cal X}^2\right],
\end{equation}
with then
\begin{equation}
L_T^{\mu\nu}=x^\mu\Pi^\nu_x-x^\nu\Pi^\mu_x+p^\mu\Pi^\nu_p-x^\nu\Pi^\mu_p=
x^\mu{\cal P}^\nu-x^\nu{\cal P}^\mu+p^\mu{\cal X}^\nu-p^\nu{\cal X}^\mu=
L_{\rm orbital}^{\mu\nu}+p^\mu{\cal X}^\nu-p^\nu{\cal X}^\mu.
\end{equation}
In this case the sector $(p^\mu,\Pi^\mu_p)$ plays the r\^ole of internal degrees of freedom,
with $(x^\mu,\Pi^\mu_x)$ that of the spacetime ones. Diagonalisation is best achieved in the
conjugate momentum basis of $(\Pi^\mu_x,\Pi^\mu_p)=({\cal P}^\mu,{\cal X}^\mu)$ eigenstates.
In that case no negative norm states arise in Hilbert space, but the Hamiltonian spectrum is
no longer bounded below, since both ${\cal P}^2$ and ${\cal X}^2$ may take arbitrarily
negative eigenvalues, albeit all in a manifestly Lorentz covariant manner. Note that
any of the energy-momentum eigenstates of fixed ${\cal P}^\mu$ is infinitely non-countably
degenerate in the spectrum of ${\cal X}^\mu$ for a fixed value of ${\cal X}^2$ (when
$\Pi_\theta\ne 0$, this infinite degeneracy for a fixed ${\cal P}^\mu$ remains discrete).
The internal spin representations are those of tensors of all orders characterised by the operator
$(p^\mu{\cal X}^\nu-p^\nu{\cal X}^\mu)$ with $(p^\mu,{\cal X}^\mu)$ spanning the Heisenberg algebra
$\left[p^\mu,{\cal X}^\nu\right]=i\hbar\eta^{\mu\nu}$.

\section{World-Line Quantisation and Physical States}
\label{Sec4}

Coupling the system to a world-line metric in a diffeomorphic invariant way is simple enough.
As explained in \cite{Gov2}, this is achieved through the action
\begin{equation}
S[x^\mu,p^\mu,\theta;e]=\int d\tau\left[\frac{1}{e}L+e\Lambda\right],
\end{equation}
where the dimensionless degree of freedom $e(\tau)$ -- actually a Lagrange multiplier -- is the world-line einbein
(using $|e(\tau)|$ rather than $e(\tau) $ in the above action implies invariance also
under orientation reversing world-line diffeomorphisms), while $L$ is the Lagrange function in (\ref{eq:action1}).
Here, $\Lambda$ is a real constant of the same physical dimension as $L$ (and $H$), playing the r\^ole
of a cosmological constant on the world-line \cite{Gov2}.
 
Applying the usual constraint analysis \cite{Govbook} the Hamiltonian formulation of the system is given by
\begin{equation}
S=\int d\tau\left\{\dot{x}\cdot\Pi_x+\dot{p}\cdot\Pi_p+\dot{\theta}\Pi_\theta
\,-\,e\left(H-\Lambda\right)\right\},
\end{equation}
$H$ being the Hamiltonian of which the quantisation has been considered above. It should be
obvious that the present dynamics preserves all the previous global symmetries associated
to the quaplectic group, whereas invariance under $\tau$ translations has been promoted to
a local gauge invariance on the world-line. The first class constraint generating the latter
symmetry is the condition
\begin{equation}
\Phi=H-\Lambda=0,\qquad H=\Lambda,
\end{equation}
which defines the space of physical states, {\it i.e.\/}, states invariant under the world-line
diffeomorphism symmetry group. Since all previously identified Noether charges
commute with $H$, indeed quantum physical states fall into representations of the quaplectic group.
For physical consistency, the choice of cosmological constant $\Lambda$ must be such that
none of these physical states be of negative norm.

In any $\Pi_\theta$ superselection sector such that $\Pi_\theta=\hbar(n+\lambda)\ne 0$, the $\hat{H}$ eigenspectrum
is given by
\begin{equation}
H\ :\qquad \frac{1}{2}N_0\frac{\lambda_0}{\alpha\kappa_0}\hbar^2(n+\lambda)^2\,+\,
\hbar N_0\frac{|\lambda_0\hbar(n+\lambda)|}{\sqrt{\kappa_0}}
\left(2N+D\right),
\end{equation}
while we know that no negative norm states are present only in the lowest level $N=0$.
It should be clear that given a choice of values for the parameters $\kappa_0$,
$\lambda_0$ and $\alpha$, there always exists a range of values for $\Lambda$ such
that a solution to the condition $H=\Lambda$ exists only with $N=0$ and for some unique
choice of $\lambda$ and $n$. In this case the gauge invariance
constraint projects out from the physical spectrum all negative norm states.
We also know that this physical spectrum then includes states of energy-momentum
${\cal P}^\mu$ taking an arbitrary value, and such that these are organised
into representations of the Poincar\'e group all of vanishing spin. However,
the mass spectrum thereby obtained is not at all constrained, with the invariant
mass-squared value $(Mc)^2=-{\cal P}^2$ taking on all real values, hence
corresponding to a continuous mass spectrum of massive states, a massless state, and
a continuum of space-like tachyonic states.

In the $\Pi_\theta$ superselection sector with $\Pi_\theta=0$, namely when $n=0$ with the
choice $\lambda=0$, no negative norm states arise,
and the gauge invariance constraint then reduces to
\begin{equation}
-(Mc)^2={\cal P}^2=\frac{2}{N_0}\Lambda-\frac{1}{\kappa_0}{\cal X}^2.
\end{equation}
However since no restriction applies to the spectra of ${\cal X}^\mu$ and ${\cal X}^2$ eigenvalues,
once again a continuum mass spectrum including a space-like tachyonic branch is obtained, with
a spin content associated to the internal degrees of freedom which spans all possible tensor
representations of the corresponding little groups according to whether the states are time-,
light- or space-like.

\section{Conclusion}
\label{Sec5}

This paper solved the world-line quantisation of a simple particle-like system realising
the quaplectic symmetry inherent to the Born-Green reciprocity principle under the exchange of
spacetime coordinates $x^\mu=(ct,\vec{x}\,)$ and their conjugate variables $p^\mu=(E/c,\vec{p}\,)$
for a Minkowski spacetime geometry. By considering free motion on the associated Weyl-Heisenberg group,
an action invariant under the full quaplectic group $Q(D-1,1)\cong U(D-1,1)\ltimes H(D)$,
which includes the ordinary Poincar\'e group as a subgroup, was constructed, and coupled to a
world-line geometry in a diffeomorphic invariant manner. Within the quaplectic realisation of the Born-Green
reciprocity principle, this is indeed the simplest generalisation possible of the similar
construction for the relativistic scalar particle, in which case the symmetry group over which free
motion is considered is the abelian spacetime translation group. An intriguing outcome of
the construction is that the spectrum of the Hamiltonian is organised into Landau-like levels.

The coupling to the world-line geometry involves a constant parameter akin to a world-line
cosmological constant. Even though the space of quantum states
includes negative norm ones for a manifestly spacetime covariant quantisation, the structure
of the diffeomorphic gauge invariance constraint is such that for an appropriate choice of
the cosmological constant none of the negative norm states belongs to the physical spectrum,
in which case the latter are all of zero spin. However, the physical spectrum is infinitely
degenerate in the conserved total energy-momentum of the system -- the symmetry associated
to spacetime translations --, with no further restriction on the corresponding Lorentz invariant
mass-squared quantity taking all possible real values. Consequently, the physical spectrum consists 
of relativistic scalar particles but with a continuous mass spectrum including a space-like tachyonic
branch. Given the general scheme of the quaplectic group as a symmetry for reciprocal
invariance in phase space, which includes the Poincar\'e algebra as a subgroup, the appearance of
such a continuous mass spectrum certainly seems to be consistent with the conclusions of
O'Raifeartaigh's theorem \cite{ORaif}.

A continuous mass spectrum is not an appealing feature from a physics point of view. However,
it is quite noteworthy that by tuning the value of a simple constant parameter in the action, namely
the cosmological constant term $\Lambda$, it is possible to project out negative norm states from
the physical spectrum, even though the appearance of tachyonic states cannot be avoided. In the
case of the ordinary relativistic spinless particle, no negative norm states are possible but this time
the cosmological term, which sets the mass of the particle, determines whether the state is
time-, light- or space-like. In the context of string theory in the critical spacetime dimension,
in effect the world-sheet cosmological constant is set to zero because of world-sheet conformal invariance, and
indeed all strictly negative norm states are once again projected out from the physical spectrum while
space-like tachyonic states are not necessarily absent on account of world-sheet reparametrisation
invariance alone. This raises the question whether for two-dimensional quantum gravity there
could exist also nonvanishing values of the cosmological constant which would once again project out
the negative norm states from the physical spectrum.

One would like to identify possible restrictions on the construction such that both
a continuous mass spectrum is avoided and its tachyonic branch excluded. A first naive attempt
would be to identify the conserved energy-momentum of the system with the dual conjugate
coordinates $p^\mu$, and then in accord with the reciprocity principle, do likewise for the
spacetime coordinates $x^\mu$ and the conserved quantities associated to translations in
the conjugate coordinates. Namely one may wish to impose the following extraneous
second-class constraints,
\begin{equation}
{\cal P}^\mu=p^\mu,\qquad
{\cal X}^\mu=x^\mu.
\end{equation}
However, an interpretation in terms of relativistic states of specific invariant mass and spin
requires consistency between the conservation of these quantities and the equations of motion,
namely
\begin{equation}
\dot{x}^\mu(\tau)=0,\qquad
\dot{p}^\mu(\tau)=0.
\end{equation}
Such restrictions thus prove to be too stringent.

One may wonder whether, by rendering the construction world-line
supersymmetric, some of the problems could not be avoided. One could
indeed hope that a tachyonic branch may be excluded with the help of
some spacetime supersymmetric projection of the physical spectrum -- as
happens with the GSO projection of fermionic strings. However, it is not
clear how a continuous mass spectrum, albeit time- and light-like only,
could be avoided. One possibility would be to identify a way of
compactifying the conserved energy-momentum ${\cal P}^\mu$, and hence
also ${\cal X}^\mu$, on account of Born-Green reciprocity. Since, when
$\Pi_\theta\ne 0$, these two quantities do not commute, this would imply
not only a discrete mass spectrum, but a finite one as well. Such a full
compactification would then render finite the volume of the phase space
associated to the conserved quantities $({\cal X}^\mu,{\cal P}^\mu)$,
namely the generators of the Weyl-Heisenberg subgroup $H(D)$ of the
full quaplectic symmetry $Q(D-1,1)\cong U(D-1,1)\ltimes H(D)$, of
the reciprocally invariant system studied in this work.

\section*{Acknowledgements}

J.~G. acknowledges the Institute of Theoretical Physics for an Invited Research Staff position at the
University of Stellenbosch (Republic of South Africa).
He is most grateful to Profs. Hendrik Geyer and Frederik Scholtz, and the School of Physics
for their warm and generous hospitality during his sabbatical leave, and for financial support.
His stay in South Africa is also supported in part by the Belgian National
Fund for Scientific Research (F.N.R.S.) through a travel grant.

J.~G. acknowledges the Abdus Salam International Centre for Theoretical
Physics (ICTP, Trieste, Italy) Visiting Scholar Programme
in support of a Visiting Professorship at the UNESCO-ICMPA (Republic of Benin).

This work was initiated during visits at the University of Tasmania. J.~G. wishes to thank
Prof. P. D. Jarvis and his colleagues of the School of Mathematics and Physics for their warm
and welcoming hospitality, the Australian Research Council for financial support for the first visit
in December 2003, and the Jane Franklin Hall College of the University of Tasmania for generous accommodation during
the second visit in December 2006. 

The work of J.~G. is also supported by the Institut Interuniversitaire des Sciences Nucl\'eaires and by
the Belgian Federal Office for Scientific, Technical and Cultural Affairs through
the Interuniversity Attraction Poles (IAP) P6/11.

\section*{Appendix}

In this Appendix the choice of action (\ref{eq:action1}) used in Sec. \ref{Sec2} is motivated
by considering free motion on the Weyl-Heisenberg group associated to the $D$ dimensional
Minkowski spacetime coordinates $x^\mu=(ct,\vec{x})$ and their conjugate variables $p^\mu=(E/c,\vec{p}\,)$. 
Indeed, from the outset such an approach is guaranteed to lead to a system invariant
under the full quaplectic group $Q(D-1,1)\cong U(D-1,1)\ltimes H(D)$. The Weyl-Heisenberg
group is generated by hermitian operators $\hat{X}^\mu$, $\hat{P}^\mu$ and the unit
operator $\mathbb{I}$ such that
\begin{equation}
\left[\hat{X}^\mu,\hat{P}^\nu\right]=i\hbar\,\eta^{\mu\nu}\,\mathbb{I}.
\label{eq:Heisenberg}
\end{equation}
The general Weyl-Heisenberg group element is thus parametrised according to
\begin{equation}
g(\theta,x^\mu,p^\mu)=
e^{i\theta\mathbb{I}+\frac{i}{\hbar}p^\mu\hat{X}_\mu-\frac{i}{\hbar}x^\mu\hat{P}_\mu},
\end{equation}
where the angular variable $\theta$ takes its values, say, in the interval $0\le\theta\le 2\pi$.

In the case of a general Lie group $G$ of elements $g$, the $G$ invariant line element on the
group manifold is given in the form, at least at a formal level,
\begin{equation}
ds^2=-{\rm Tr}\,\left(g^{-1}dg\right)^2,
\end{equation}
up to some normalisation factor, and the necessity of a proper definition of the trace operation.
In turn the action for free motion on such a manifold is of the form, again up to normalisation
\begin{equation}
S[g]\propto -\int dt\,{\rm Tr}\,\left(g^{-1}\frac{dg}{dt}\right)^2.
\end{equation}

In the case of the above Weyl-Heisenberg group elements $g(\theta,x^\mu,p^\mu)$ a direct calculation
finds
\begin{equation}
-{\rm Tr}\,\left(g^{-1}dg\right)^2=
\frac{1}{\hbar^2}\left[dx^\mu dx^\nu\,{\rm Tr}\,\hat{P}_\mu\,\hat{P}_\nu\ +\
dp^\mu dp^\nu\,{\rm Tr}\,\hat{X}_\mu\,\hat{X}_\nu\right]\ +\
\left[d\theta-\frac{1}{2\hbar}\left(dx^\mu\,p_\mu\,-\,x^\mu dp_\mu\right)\right]^2\,
{\rm Tr}\,\mathbb{I},
\end{equation}
in which it is assumed that the definition of the trace operation is such that the
operators $\hat{X}^\mu$, $\hat{P}^\mu$, $\hat{X}^\mu\hat{P}^\nu$ and $\hat{P}^\mu\hat{X}^\nu$
are of vanishing trace. Introducing then a regularised definition of the trace operation in
the case of the infinite dimensional representation of the Heisenberg algebra (\ref{eq:Heisenberg}),
it follows that
\begin{equation}
-{\rm Tr}\,\left(g^{-1}dg\right)^2\propto
\frac{1}{2}\left[\,dx^\mu dx_\mu\,+\,\mu_0 dp^\mu dp_\mu\right]+
\frac{1}{2}\mu_1\left[d\theta-\frac{1}{2\hbar}\left(dx^\mu p_\mu-x^\mu dp_\mu\right)\right]^2,
\end{equation}
where $\mu_0>0$ and $\mu_1\ge 0$ are regularisation dependent normalisation factors of the appropriate
physical dimension. This latter expression provides the choice of action considered in Sect. \ref{Sec2}.

An alternative realisation of the Weyl-Heisenberg group is provided by the following $(2D+2)\times(2D+2)$ real matrices~\cite{Low2005b},
\begin{equation}
H(\theta,x^\mu,p_\mu)=\left(\begin{array}{c c c c}
\mathbb{I}_D & 0 & 0 & p_\mu \\
0 & \mathbb{I}_D & 0 & x^\mu \\
-x^\mu & p_\mu & 1 & 2\theta \\
0 & 0 & 0 & 1
\end{array}\right),\quad {\rm with}\quad
dH(\theta,x^\mu,p_\mu)=\left(\begin{array}{c c c c}
0 & 0 & 0 & dp_\mu \\
0 & 0 & 0 & dx^\mu \\
-dx^\mu & dp_\mu & 0 & 2d\theta \\
0 & 0 & 0 & 0
\end{array}\right),
\end{equation}
where $x^\mu,p_\mu\in\mathbb{R}$ and $0\le\theta<2\pi$. The line element
$d\ell^2=\frac{1}{4}{\rm Tr}\,(H^{-1}dH)^2$ is thus
\begin{equation}
d\ell^2=\frac{1}{2}dx^\mu dx_\mu+\frac{1}{2} dp^\mu dp_\mu+
\frac{1}{2}\left[d\theta-\left(dx^\mu p_\mu - x^\mu dp_\mu\right)\right]^2,
\end{equation}
in which the Minkowski metric is used to raise or lower indices where necessary.
Up to scalings of $\theta$ and $p_\mu$ relative to $x^\mu$ by appropriate dimensional constants,
this expression reproduces the required line element for our world-line action.

\end{document}